\documentclass[aip,apl,amsmath,amssymb,reprint]{revtex4-1}

\usepackage{graphicx}
\usepackage{dcolumn,colortbl}
\usepackage{bm}

\begin{document}

\newcommand{\bt}[1]{{\color{red}$\clubsuit$#1}}
\newcommand{\pb}[1]{{\color{blue}$\spadesuit$#1}}

\preprint{APL/xxx}

\title{Structural properties and Thermodynamics of Hafnium sub-oxides in RRAM}

\author{Philippe Blaise}
\email{philippe.blaise@cea.fr}
\affiliation{Univ. Grenoble Alpes, F-38000 Grenoble France.}
\affiliation{CEA, LETI, MINATEC Campus, F-38054 Grenoble, France.}
\email{traorehade@yahoo.fr}
\author{Boubacar Traore}
\affiliation{Univ. Grenoble Alpes, F-38000 Grenoble France.}
\affiliation{CEA, LETI, MINATEC Campus, F-38054 Grenoble, France.}
\affiliation{Fondation Nanoscience, 25 rue des Martyrs, 38000 Grenoble, France}

\date{\today}

\begin{abstract}
We study the structural and electronic properties of various hafnium sub-oxides Hf$_z$O from $z = 9$ to $z = 0.5$, by {\it ab initio} simulation using Density Functional Theory. The stability of these sub-oxides is studied against monoclinic HfO$_2$. The progressive oxidation of a given Hf$_z$O is also envisaged toward stoichiometric HfO$_2$. The analogy with a conductive region of electrons inside a HfO$_2$ matrix is discussed within the context of Oxide-based Resistive Random Access Memories (OxRRAM) devices which employ hafnium dioxide as an insulator. 
\end{abstract}

\pacs{XXX}

\maketitle

%%%%%%%%%%%

Hafnium dioxide and its alloys have definitely emerged as the standard dielectric for SiO$_2$ replacement in ultra-scaled CMOS technology\cite{Wilk_2001,Mistry_2007}. In this context, oxygen related defects at a relatively low concentration have been widely studied due to their importance for good threshold voltage control, tunneling current interpretation due to traps and reliability issues\cite{McIntyre_2007}. In parallel, resistive random access memories RRAM have attracted a lot of attention for their development and understanding\cite{Jeong_2012}. Among the different emerging technologies, RRAM based on a metal oxide or OxRRAM\cite{Beck_2000,Baek_2004,Wong_2012} are one of the best candidates to pursue the non volatile memory scaling. Generally speaking an OxRRAM made with HfO$_2$, (directly compatible with CMOS integration), shows a quasi reversible switching between a low and a high resistance state due to the resistance modulation of a conductive filament of nanometric size\cite{Bersuker_2011,Wong_2012}. The existence of this filament, which has never been directly identified, is mainly supported by electrical measurements and modeling \cite{Bersuker_2011,Procel_2013}. As far as the filament hypothesis seems to be relevant, all these models share the hypothesis that is based on oxygen movements between one of the electrodes and the dielectric that leads to the formation of sub-oxides which we denote HfO$_x$ or Hf$_z$O inside the host oxide. Nevertheless Hafnium sub-oxides are not well characterized except for the hafnium-rich compounds of type Hf$_z$O with $z \geq 2$, where the sub-lattice spanned by hafnium atoms retains its original hexagonal structure up to the oxygen solubility limit in pure hcp-Hf\cite{Burton2012151}. More recently, a semi-metal with a tetragonal Hf$_2$O$_3$ structure has been predicted by {\it ab initio} simulation\cite{Xue_2013}. Moreover, a cousin zirconium-oxide of the hexagonal form ZrO has been predicted as even more stable than the tetragonal Zr$_2$O$_3$\cite{Puchala_2013}.

Therefore, we propose ourselves to study by {\it ab initio} methods the known hafnium sub-oxides and recently predicted forms. Our simulation results will be useful in order to sustain the filament hypothesis for OxRRAM technology and the related physical characterization experiments. 

%%%%%%%%%%%

Our {\it ab initio} calculations are performed using Siesta\cite{siesta_web,siesta} which implements a generalized gradient approximation (GGA) of density functional theory (DFT) \cite{Book_parr_Yang} thanks to the Perdew-Burke-Ernzerhof (PBE)\cite{Perdew_1996} parametrization. Siesta employs a linear combination of numerical atomic orbitals as the basis set to describe the wave functions \cite{siesta}. A double zeta plus polarization (DZP) basis set with an energy shift of $50$ meV is used in all our calculations. Troullier-Martins pseudopotentials \cite{Troullier_1991} were derived with the configurations O $2s^{2}2p^{4}$ and Hf $5d^{2}6s^{2}$, including a non-linear core correction and a relativistic correction for Hafnium. The structural parameters and atomic positions were optimized using a conjugate gradient scheme, until the maximum residual forces and stresses were less than $0.02$ eV/$\mbox{\AA}$ and $200$ MPa, respectively. To obtain realistic values for the band gap energy, we employed the half-occupation {\it ab initio} technique GGA-$1/2$ \cite{Luiz_2008} to correct for the self-energy of the oxygen anion. This technique is convenient for extended systems producing band gaps which are in good agreement with experiment and other band gap correction schemes at a negligible extra computational effort \cite{Ribeiro_2009}.

%%%%%%%%%%%

The different Hf$_z$O oxides formula we used are summarized in table \ref{table.1_HfOx_structures} with their respective space groups and symmetries. Hafnium metal is known to be hexagonal and presents an unusual high oxygen solubility of 28\%\cite{Burton2012151}. This high solubility is due to the insertion of oxygen atoms inside the octahedral interstices all along specific planes accompanied by a moderate expansion of the c axis of Hafnium. This allows it to retain the hexagonal symmetry down to $z = 2$. For $z = 1$ ie. HfO(a), we employed the hexagonal h-ZrO structure proposed by Puchala\cite{Puchala_2013} while for Hf$_2$O$_3$ we used the tetragonal structure obtained by Xue\cite{Xue_2013}. In order to reach the full stoichiometry of HfO$_2$, we used several other structures: HfO(b) which is obtained by removing an oxygen atom from t-Hf$_2$O$_3$, Hf$_4$O$_7$(a) obtained by inserting a supplementary oxygen inside t-Hf$_2$O$_3$, Hf$_4$O$_7$(b) by removing an oxygen atom from a monoclinic supercell of HfO$_2$ and Hf$_{32}$O$_{63}$ which is a neutral oxygen vacancy inside a 2x2x2 supercell of monoclinic HfO$_2$. These last structures are not unique but representative of the numerous possibilities that we can expect from these intermediate compositions, (see the discussion below).
  
\renewcommand{\arraystretch}{1.4}
\begin{table}[!ht]
\begin{center}
\begin{tabular}{|l|l|l|c|}
 \hline
 Formula            &  Symmetry     &  Space Group         &  Ref.                  \\ \hline
 Hf                 &  Hexagonal    &  P6$_3/$mmc (\#194)  &  \cite{Russell_1953}   \\ \hline
 Hf$_9$O            &  Hexagonal    &  P$\bar{3}$ (\#147)  &                        \\ \hline
 Hf$_6$O            &  Hexagonal    &  R$\bar{3}$ (\#148)  &  \cite{Burton2012151}  \\ \hline
 Hf$_4$O            &  Hexagonal    &  P$\bar{3}$ (\#147)  &                        \\ \hline
 Hf$_3$O            &  Hexagonal    &  P$\bar{3}$1c (\#163)&  \cite{Burton2012151}  \\ \hline
 Hf$_2$O            &  Hexagonal    &  P$\bar{3}$1m (\#162)&  \cite{Burton2012151}  \\ \hline
 HfO(a)             &  Hexagonal    &  P$\bar{6}$2m (\#189)&  \cite{Puchala_2013}   \\ \hline
 HfO(b)             &  Tetragonal   &  P4nmm (\#129)       &                        \\ \hline
 Hf$_2$O$_3$        &  Tetragonal   &  P$\bar{4}$m2 (\#115)&  \cite{Xue_2013}       \\ \hline
 Hf$_4$O$_7$(a)     &  Tetragonal   &  P$\bar{4}$2m (\#111)&                        \\ \hline
 Hf$_4$O$_7$(b)     &  Monoclinic   &  P2$_1/$c (\#14)     &                        \\ \hline
 Hf$_{32}$O$_{63}$  &  Monoclinic   &  P2$_1/$c (\#14)     &                        \\ \hline 
 HfO$_2$            &  Monoclinic   &  P2$_1/$c (\#14)     &  \cite{Lowther_1999}   \\ \hline    
\end{tabular}
\end{center}
\caption{Hf$_z$O crystal structures employed with their respective symmetries and space groups.}
\label{table.1_HfOx_structures}
\end{table}

All along the path of oxidation of Hafnium it is essential for our study to extract the essential characteristics of the Hf$_z$O sub-oxides. In table \ref{table.2_Structural_Properties} we give the results we obtained in DFT/GGA for: volume of formula unit per Hf atom, its relative increase from the initial hexagonal Hafnium configuration, the mean coordination numbers of Hafnium and Oxygen atoms, the mean bond lengths of Hf$-$Hf and Hf$-$O bonds. Then in table \ref{table.3_Physical_Properties} we give: the formation enthalpies per mole of oxygen obtained in GGA (see Eq. \ref{eq.DH}), the electronic band gap estimated with the half-occupation technique in GGA-$1/2$ and the mean Bader charges for Hf and O atoms. 

\renewcommand{\arraystretch}{1.4}
\begin{table}[!ht]
\begin{center}
\begin{tabular}{|l|l|l|r|r|r|r|}
 \hline
 Formula            & V$/$Hf        & \%inc.     & \#Hf & \#O & $\overline{\text{Hf-Hf}}$ & $\overline{\text{O-Hf}}$ \\ \hline
 Hf                 &  22.8(22.3)   &   +0.0     & 12.   & 0. & 3.2 &     \\ \hline
 Hf$_9$O            &  23.0         &   +0.9     & 12.7  & 6. & 3.2 & 2.3 \\ \hline
 Hf$_6$O            &  23.1         &   +1.3     & 13.   & 6. & 3.2 & 2.3 \\ \hline
 Hf$_4$O            &  23.2         &   +1.8     & 13.5  & 6. & 3.2 & 2.3 \\ \hline
 Hf$_3$O            &  23.3         &   +2.2     & 14.   & 6. & 3.2 & 2.3 \\ \hline
 Hf$_2$O            &  23.4         &   +2.6     & 15.   & 6. & 3.2 & 2.3 \\ \hline
 HfO(a)             &  25.4         &  +11.4     & 17.   & 5. & 3.3 & 2.2 \\ \hline
 HfO(b)             &  25.4         &  +11.4     & 17.   & 5. & 3.3 & 2.3 \\ \hline
 Hf$_2$O$_3$        &  28.0         &  +22.8     & 19.   & 4.7& 3.5 & 2.2 \\ \hline
 Hf$_4$O$_7$(a)     &  29.7         &  +30.3     & 18.   & 4.6& 3.5 & 2.3 \\ \hline
 Hf$_4$O$_7$(b)     &  35.6         &  +56.1     & 13.   & 3.4& 3.5 & 2.2 \\ \hline
 Hf$_{32}$O$_{63}$  &  35.6         &  +56.1     & 13.9  & 3.5& 3.5 & 2.2 \\ \hline 
 HfO$_2$            &  35.6(34.6)   &  +56.1     & 14.   & 3.5& 3.5 & 2.2 \\ \hline    
\end{tabular}
\end{center}
\caption{Hf$_z$O structural properties: volume in $\AA^3$ per Hafnium atom with experimental values inside parentheses, relative volume increase in \% per Hf atom, \# of Hf and O first neighbors, mean atomic distance of Hf$-$Hf and O$-$Hf bonds in $\AA$.}
\label{table.2_Structural_Properties}
\end{table}

As can be seen in tables \ref{table.1_HfOx_structures} and \ref{table.2_Structural_Properties}, below $z=2$ the phase change from hexagonal to tetragonal to monoclinic symmetries is accompanied by a large volume increase due to oxygen insertion. At the same time, the O coordination number drops from 6 in octahedral position related to the high oxygen solubility in Hf to 3 and 4 for monoclinic HfO$_2$. During this volume increase from hexagonal to monoclinic, Hf$-$Hf bond lengths are stretched from $3.2$ to $3.5$ $\AA$ while O$-$Hf bond lengths are kept almost constant at 2.2$-$2.3 $\AA$. 

\renewcommand{\arraystretch}{1.4}
\begin{table}[!ht]
\begin{center}
\begin{tabular}{|l|c|c|c|c|}
 \hline
 Formula            &  $\Delta$H    & $\,$ E$_g\,$ &  $\overline{\text{Q}}_{Hf}$ &  $\overline{\text{Q}}_O$  \\ \hline
 Hf                 &  0            &  0.0   &  0.0  &  0.0  \\ \hline
 Hf$_9$O            &  -561   &  0.0   &  0.2  & -1.8  \\ \hline
 Hf$_6$O            &  -566   &  0.0   &  0.3  & -1.8  \\ \hline
 Hf$_4$O            &  -552   &  0.0   &  0.5  & -1.8  \\ \hline
 Hf$_3$O            &  -546   &  0.0   &  0.6  & -1.8  \\ \hline
 Hf$_2$O            &  -532   &  0.0   &  0.9  & -1.8  \\ \hline
 HfO(a)             &  -521   &  0.4   &  1.6  & -1.6  \\ \hline
 HfO(b)             &  -498   &  0.0   &  1.5  & -1.5  \\ \hline
 Hf$_2$O$_3$        &  -513 &  0.0   &  2.2  & -1.5  \\ \hline
 Hf$_4$O$_7$(a)     &  -491 &  0.4   &  2.4  & -1.4  \\ \hline
 Hf$_4$O$_7$(b)     &  -500 &  1.1   &  2.4  & -1.4  \\ \hline
 Hf$_{32}$O$_{63}$  &  -513 &  5.8   &  2.7  & -1.4  \\ \hline 
 HfO$_2$            &  -515 &  5.8   &  2.7  & -1.4  \\ \hline    
\end{tabular}
\end{center}
\caption{Hf$_z$O physical properties: formation enthalpy in kJ/mol of oxygen obtained in GGA, estimated band gap in eV (GGA-$1/2$), mean Bader charge for Hf and O in $|e|$.}
\label{table.3_Physical_Properties}
\end{table}

The formation enthalpies are estimated at T$=$0 K and P$=$0 Bar by using the hexagonal Hafnium metal and molecular oxygen as reference states. This corresponds to the following formula where temperature and pressure effects are usually neglected for solid states allowing us to use directly the internal energies we obtained in GGA:
\begin{equation}
\Delta H_{f}(\text{Hf$_z$O}) = E_{tot}(\text{Hf$_z$O}) - z.E_{tot}(\text{hcp-Hf}) - \frac{1}{2}.E_{tot}(\text{O$_2$})
\label{eq.DH}
\end{equation}
Eq. \ref{eq.DH} corresponds to the formation enthalpy per mole of oxygen employed to create an Hf$_z$O alloy. This formulation permits to gauge enthalpies of sub-oxides with different stoichiometries with the help of Tab. \ref{table.3_Physical_Properties}. By comparing with the experimental value of $\Delta$H $=$ -1144.7 kJ$/$mol of HfO$_2$, we recover the well known underbinding effect due to partial cancellation of errors in GGA: $-1144.7/2. = -572$ kJ$/$mol of oxygen (exp.) vs -515 kJ$/$mol (GGA).

We now turn to the basic electronic properties of the sub-oxides. Table \ref{table.2_Structural_Properties} and \ref{table.3_Physical_Properties} show that the mean Bader charge increase for Hf atoms is directly correlated to the Hf$_z$O volume increase, to a lower number of first neighbor oxygen atoms and to the band gap opening. For the electronic band gap the situation seems however to be more complex: the original bang gap of m-HfO$_2$ at 5.8 eV is preserved only for a small concentration of oxygen vacancies (at a few \%) but drops considerably to 0.4 eV for Hf$_4$O$_7$(a) and becomes null for Hf$_2$O$_3$ which is a semi-metal\cite{Xue_2013}. Then it increases again at 0.4 eV for hexagonal HfO(a) to vanish completely for the remaining compositions at x $\geqslant$ 2. This seems to be partially related to the Hf$-$Hf bond length which has to be sufficiently large and $\geqslant$ 3.4 $\AA$ to insure some insulating properties. For HfO(a) we also do believe that the rather specific HfO hexagonal symmetry is at the origin of its slight gap increase. Note that this mixture of (semi-)metallic and semiconductor characteristics could also be probably related to the resistance variability in OxRRAM studies\cite{Fantini_2013}. 

These calculated data constitute a starting point to describe the essential thermodynamics of HfO$_2$-based OxRRAM. In OxRRAM technology a forming step is usually required to render the initially insulating HfO$_2$ conductive, allowing electrons to flow through a Hafnium-rich conductive path or filament\cite{Xue_2014}. By neglecting the interface effects (i.e. for a sufficiently large filament), we can envisage the stability of a pure Hf filament in equilibrium with HfO$_2$ against oxidation by considering the following reaction:
\begin{equation}
\text{Hf} + \frac{x}{2-x}.\text{HfO$_2$} \longrightarrow  \frac{2}{2-x}.\text{HfO$_x$} \;\; 0 \leqslant x < 2
\label{eq.HfOx}
\end{equation}   
%%%%%%%%%%%

\begin{figure}[!h]
\begin{center}
\includegraphics[scale=0.31]{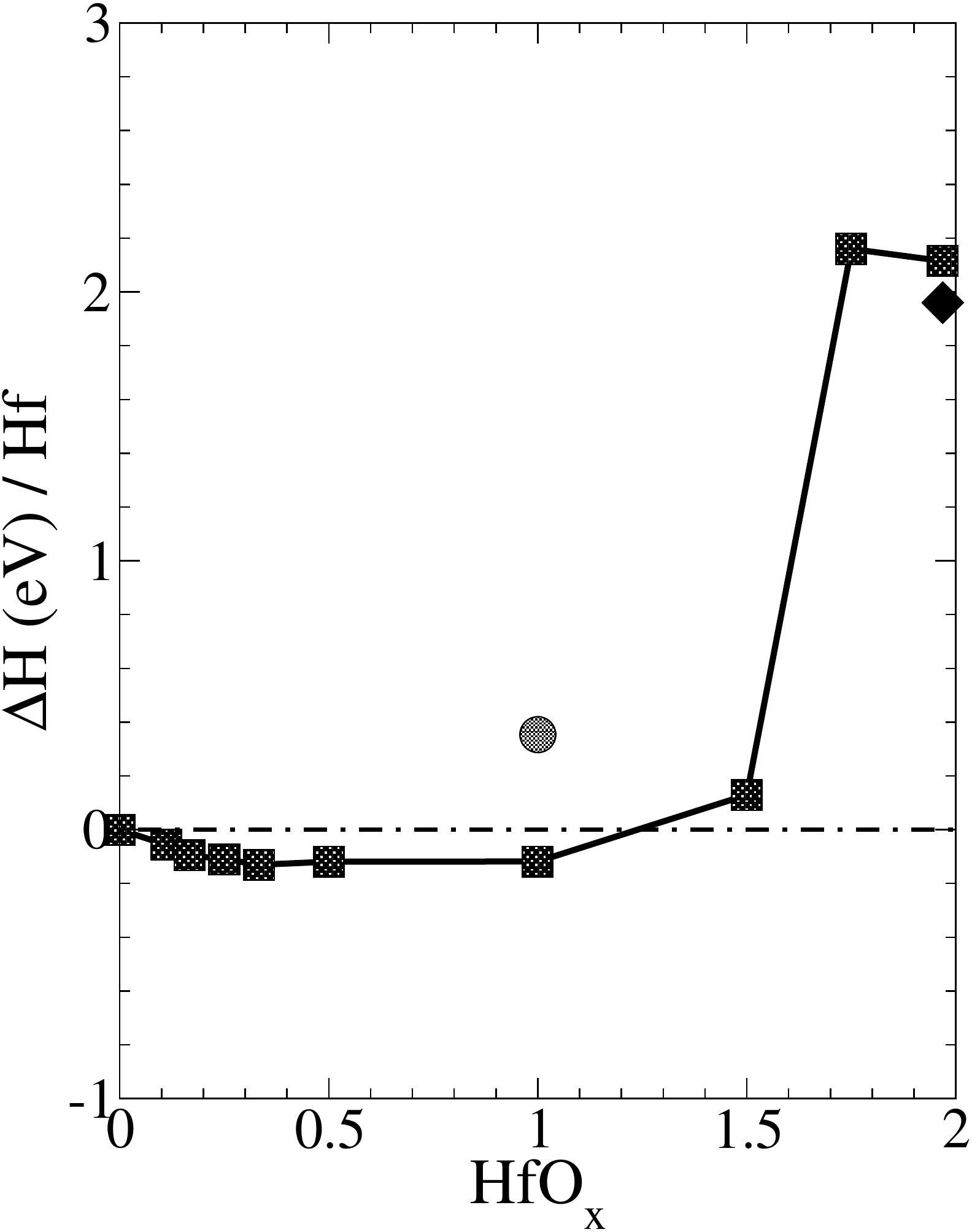}
\caption{HfO$_x$ stability against pure Hf and HfO$_2$ in infinite proportion. The black curve is derived using eq. \ref{eq.HfOx} and table \ref{table.3_Physical_Properties}. For $x=1$ the upper point is for HfO(b) structure. For $x$ close to $2$ we also show the energy obtained using data from.\cite{Zheng_2007}}
\label{fig.HfOx}
\end{center}
\end{figure}

The basic idea behind equilibrium Eq. \ref{eq.HfOx} is to consider the pure Hafnium filament as a limit of an Hf-alloy with oxygen atoms shared with the surrounding HfO$_2$. The corresponding enthalpy curve for equilibrium Eq. \ref{eq.HfOx} is shown in Fig. \ref{fig.HfOx}. At $x=0$ the pure Hf limit serves as a reference energy $\Delta H=0$, indicating that the O chemical potential is assumed to be the one in an Hf-rich environment. Therefore, for $x\rightarrow2$ the corresponding enthalpy tends exactly toward two times the formation enthalpy of an oxygen vacancy in the Hafnium-rich limit. In between, there exists a stability interval for sub-oxides enriched in oxygen from $x\approx0.2$ up to $x\approx1$. This tells that after forming, an Hf-rich region in contact with HfO$_2$ tends to form an HfO$_x$ composition, with $x$ around $0.5$ or Hf$_2$O. This is in relative agreement with the results of McKenna \cite{McKenna_2014} who proposed Hf$_5$O as the most stable composition by following a different thinking, because here we consider several new structures for the Hafnium sub-oxides with notably $x=1$. 

\begin{figure}[!h]
\begin{center}
\includegraphics[scale=0.31]{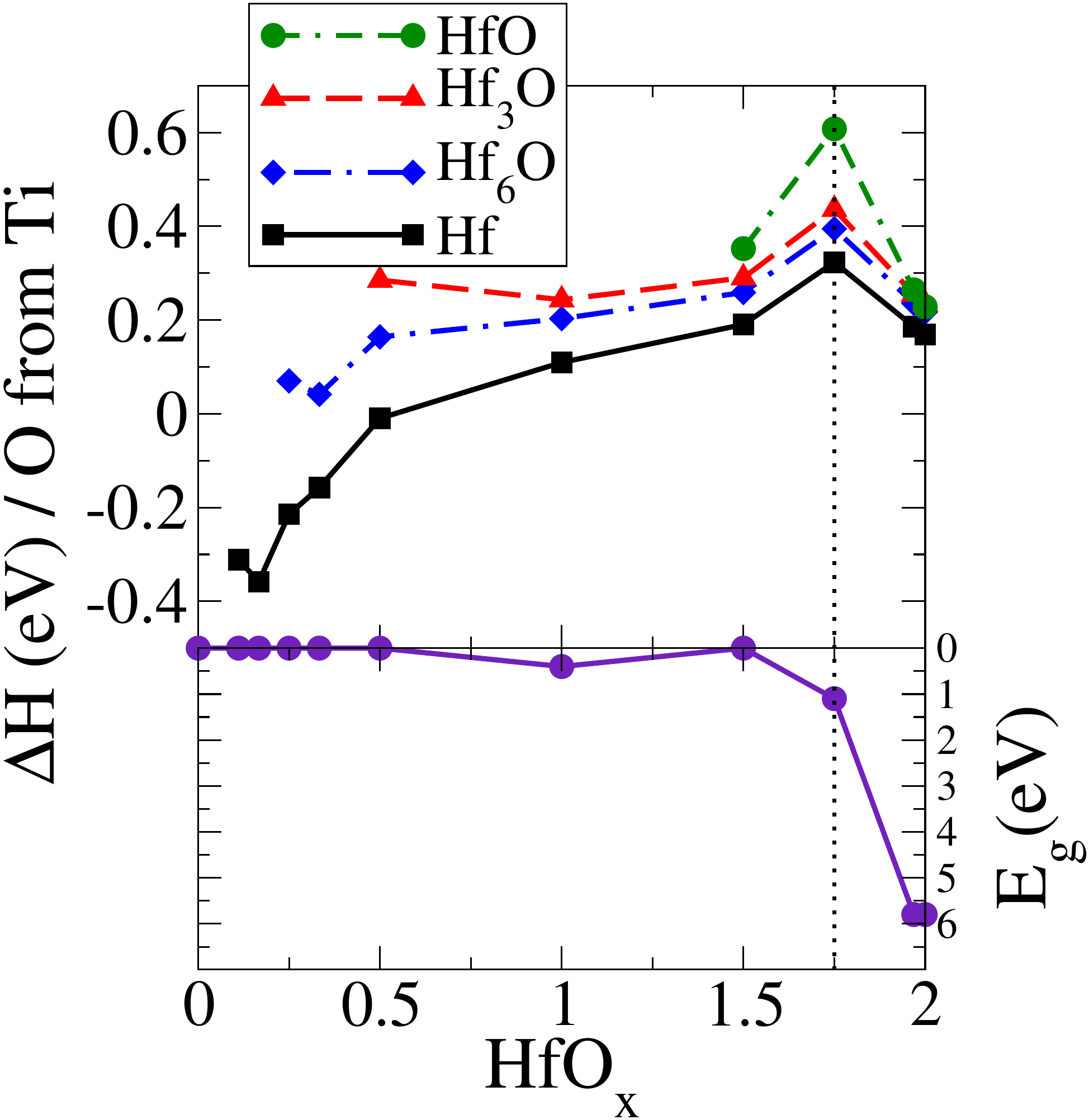}
\caption{Hf$_z$O oxidation enthalpy by oxygen atoms coming from a hcp-Ti electrode up to HfO$_2$, for various starting compositions $z = \infty, 6, 3, 1$. The lower part shows the associated forbidden band gap.}
\label{fig.HfO_ox}
\end{center}
\end{figure}

Thereafter the forming step, starting from a given composition of the filament we may wonder about the energetic of oxidation. The following equilibrium

\begin{equation}
\text{Ti$_n$:O$_i$} + \frac{1}{zx-1}.\text{Hf$_z$O} \longrightarrow  \text{Ti$_n$} + \frac{z}{zx-1}.\text{HfO$_x$} \;\; 0 \leqslant x \leqslant 2
\label{eq.TiO_HfOx}
\end{equation}      

allows us to investigate the oxidation of a filament of Hf$_z$O composition by oxygen interstitial atoms coming from a reactive electrode made of titanium in contact with the formed dielectric\cite{Chen_2010}. The $\text{Ti$_n$:O$_i$}$ and $\text{Ti$_n$}$ systems have been calculated in GGA starting from a relaxed hcp-Ti structure made of 150 atoms with one oxygen inserted in octahedral position. The obtained results for several starting compositions of Hf$_z$O namely Hf, Hf$_6$O, Hf$_3$O and HfO are shown in Fig. \ref{fig.HfO_ox}. For pure Hf and, to a lesser extent, Hf$_6$O we obtain that these compositions are stabilized toward Hf$_2$O and Hf$_3$O respectively, which definitely favors sub-oxide filaments to be stable in contact with both m-HfO$_2$ and hcp-Ti. Then, the increase of the oxygen content of a Hafnium sub-oxide in contact with Ti requires a near constant energy of 0.2 eV per oxygen atom added. The only exception occurs for the Hf$_4$O$_7$ composition which requires 0.3 eV up to 0.6 eV. Interestingly, if one follows the forbidden gap increase related to oxygen content as shown in the lower part of Fig. \ref{fig.HfO_ox}, this slight increase delimits exactly the conductive compositions from the insulating one. Therefore, we can argue that resetting an OxRRAM made of HfO$_2$ (i.e programming the device in a high resistance state) is mainly related to a bi-stable behavior of the oxidation of Hf$_z$O which can be oxidized partly as a conductor and partly as an insulator. Depending on the kinetic conditions which we do no treat here, one can imagine that the insulator fraction due to partial oxidation of a sub-oxide filament can vary a lot in position and shape leading to the intrinsic variability of the high resistive state R$_{\text{off}}$ observed experimentally for this type of OxRRAM\cite{Fantini_2013,Cabout_2013}.    

In conclusion, with the help of simulations based on density functional theory, we derived the basic properties of the most stable Hafnium sub-oxides. This allowed us to find the most stable compositions of sub-oxides in equilibrium with HfO$_2$ relative to the forming step of RRAM, but also the most stable compositions of sub-oxides in equilibrium with a reactive Titanium electrode during the RESET step. Our equilibrium calculations come from standard thermodynamics and neglect several important effects like the kinetic effects, the interface effects, which are all relevant at a nanometric scale. Nevertheless, as a first order approach, our results reveal that a conductive filament HfO$_x$ could in principle possess a large oxidation window with $0 \leqslant x \leqslant 1$. Then during the oxidation process relative to RESET, its composition would fluctuate between a conductive state with $x \leqslant 1.5$ and an insulating state with $x\approx2$. We do believe that what tells our thermodynamics calculations, is directly related to the intrinsic variability observed in HfO$_2$-based OxRRAM.     

The authors thank the Nanoscience Foundation of Grenoble (France) for its financial support.

%\nocite{*}

\bibliography{HfOx} % Produces the bibliography via BibTeX.

\end{document}